\documentclass[twocolumn]{aastex701}

\usepackage{xcolor}
\usepackage{float}
\definecolor{ochre}{rgb}{0.8, 0.47, 0.13}

\definecolor{greenish}{rgb}{0.01, 0.45, 0.33}

\begin{document}

\title{Non-conservative Mass Transfer as a Formation Channel for Gaia Black Hole Systems}

\author[0000-0002-6105-6492]{Aleksandra Olejak}
\affiliation{Max Planck Institute for Astrophysics,
Karl-Schwarzschild-Straße 1,
85748 Garching b.~München, Germany}
\email[show]{aolejak@mpa-garching.mpg.de}

\author[0000-0002-7527-5741]{Jakub Klencki} 
\affiliation{Max Planck Institute for Astrophysics,
Karl-Schwarzschild-Straße 1,
85748 Garching b.~München, Germany}
\email{}

\author[0000-0003-1817-3586]{Alejandro Vigna-Gomez} 
\affiliation{Max Planck Institute for Astrophysics,
Karl-Schwarzschild-Straße 1,
85748 Garching b.~München, Germany}
\email{}

\author[0000-0001-9336-2825]{Selma E.~de Mink}
\affiliation{Max Planck Institute for Astrophysics,
Karl-Schwarzschild-Straße 1,
85748 Garching b.~München, Germany}
\affiliation{Ludwig-Maximilians-Universität München, Geschwister-Scholl-Platz 1, 80539 München, Germany}
\email{}

\author[0000-0001-5484-4987]{Lieke van Son}
\affiliation{%
Radboud University, Faculty of Science, Huygensgebouw,
Heyendaalseweg 135, 6525 AJ Nijmegen, The Netherlands%
}
\email{}

\author[0000-0002-4914-6479]{Jakub Cehula}
\affiliation{Tartu Observatory, University of Tartu, T\~{o}ravere, 61602 Tartumaa, Estonia}
\affiliation{Astronomical Institute, Slovak Academy of Sciences, 059 60 Tatransk\'{a} Lomnica, Slovakia}
\email{}

\author[0000-0003-2340-8140]{Jakob Stegmann}
\affiliation{Max Planck Institute for Astrophysics,
Karl-Schwarzschild-Straße 1,
85748 Garching b.~München, Germany}
\email{}

\author[0000-0003-2012-5217]{Taeho Ryu}
\affiliation{JILA, University of Colorado and National Institute of Standards and Technology, 440 UCB, Boulder, 80308 CO, USA}
\affiliation{Department of Astrophysical and Planetary Sciences, 391 UCB, Boulder, 80309 CO, USA}
\affiliation{Max Planck Institute for Astrophysics,
Karl-Schwarzschild-Straße 1,
85748 Garching b.~München, Germany}
\email{}

\author[0000-0002-8717-6046]{David D. Hendriks}
\affiliation{Department of Physics, University of Surrey, Guildford, GU2 7XH, Surrey, UK}
\email{}

%\author[0000-0001-7969-1569]{Stephen Justham}
%\affiliation{Max Planck Institute for Astrophysics,
%Karl-Schwarzschild-Straße 1,
%85748 Garching b.~München, Germany}
%\email{}

\begin{abstract}

The detected Gaia systems hosting compact objects challenge standard models of binary star evolution. In particular, if the observed black hole (BH) systems evolved in isolation, they are expected to have undergone a mass transfer phase. Given their highly unequal masses, such mass transfer is dynamically unstable within standard models, leading to a stellar merger or a short-period binary. In contrast, the observed systems have much wider orbits than predicted, making their formation within conventional evolutionary frameworks difficult to reconcile.
Using detailed binary evolution calculations, we test whether non-conservative mass transfer, in which most of the mass is lost from the system carrying the specific angular momentum of the donor’s center of mass, can explain the properties of two Gaia BH systems. This mass-loss geometry differs from standard isotropic re-emission from the accretor’s vicinity.
We find that our mass-loss geometry model reproduces the orbital periods of the two Gaia BH systems remarkably well over a wide range of initial conditions, offering a plausible formation pathway. We speculate this may point to enhanced eruptive mass loss, potentially driven by high-opacity subsurface layers in the donor prior to Roche-lobe overflow, consistent with preferentially bipolar outflows observed in luminous blue variables. Alternatively, it may indicate the need for more sophisticated mass-transfer prescriptions that account for highly unequal Roche-lobe sizes, sub-synchronous rotation, and possible self-accretion. Similar mechanisms may operate in other post-mass-transfer systems facing analogous evolutionary challenges, including Gaia neutron-star and white-dwarf binaries, stripped-envelope Wolf–Rayet stars, and low-mass X-ray binaries.

\end{abstract}



\section{Introduction}

The European Space Agency's Gaia mission \citep{GAIA2016,GAIA2018,GAIA2021} has revolutionized our view of the stellar populations in the Milky Way. One of the more significant results was the identification of candidates for binary systems harboring non-accreting black holes (BHs), inferred through the orbital motion of luminous stellar companions \citep{ElBadry2023a, ElBadry2023b,Chakrabarti2023}. All so far reported BH systems: Gaia BH1, Gaia BH2, and Gaia BH3 comprise of a low mass star in a relatively wide orbit around an unseen companion with a dynamically inferred mass consistent with stellar-mass BHs. These discoveries open a new and valuable window into the demographics of BHs, complementing constraints obtained from spectroscopic surveys, X-ray binaries \citep[e.g.,][and references therein]{Cesares2014, Zdziarski2026}, the expanding catalog of gravitational-wave detections \citep{GWTC4population2025}, and the single microlensing BH candidate reported to date \citep{Sahu2022,Lam2022}.

The Milky Way is expected to host over a hundred thousand stellar-origin BHs, although the vast majority are predicted to be single.\citep{Shapiro1983,vandenHeuvel1992,Olejak2020}. The number of BHs remaining in binaries is significantly lower, possibly amounting to up to 10-20\% of the total population \citep{Lamberts2018, Renzo2019, Wiktorowicz2019, Olejak2020, Chawla2022, VignaGomez2023b}. However, this fraction depends sensitively on several uncertain factors, including the initial binary and higher-order multiplicity fractions, the contribution of stellar mergers (and the details of mass-transfer physics), and the disruption of systems during BH formation. Among the surviving binaries, most of the intrinsic population of BH-hosting systems is predicted to consist of wide, non-interacting binaries containing a BH and either another compact object or a low-mass, long-lived stellar companion \citep{Wiktorowicz2019,Breivik2017,Breivik2019,Olejak2020, Chawla2022}. From this perspective, non-interacting, relatively wide BH systems with low-mass stellar companions, such as those discovered by Gaia, were expected to be present in the Galaxy.

The real challenge emerges when considering the orbital periods of these systems: $\sim \,$186 and 1277 days \citep{Chawla2025, Nagarajan2025a}. If they formed in isolation, such separations would almost certainly require a previous phase of mass transfer within the framework of classical binary evolution models (although see \citealt{Gilkis2024} and \citealt{Kruckow2024A&A}, discussed below). Given their highly unequal mass ratios, such mass transfer is expected to be dynamically unstable \citep[e.g.,][]{Ge2015, Pavlovskii2017, Ge2020a, Ge2020b}, leading to either a stellar merger or a very tight post-common envelope (CE) binary \citep{Ivanova2013}. Therefore, the observed relatively wide orbits are inconsistent with predictions from standard binary evolution, making their formation challenging to explain. Interestingly, a similar problem has also been reported for other types of post-mass-transfer binaries, for example, the systems detected by Gaia hosting white dwarfs \citep{Yamaguchi2025} or neutron stars \citep{ElBadry2024b,Chattopadhyay2025}.

Several formation channels and stellar evolutionary models have been proposed to explain the properties of Gaia BH systems. In the case of Gaia BH3, its association with the metal-poor ED-2 stream in the Galactic halo suggests that it may have been ejected from a disrupted globular cluster, suggesting a dynamical formation scenario \citep{Pina2024, Balbinot2024}. Gaia BH3, owing to its longer orbital period and low metallicity, may have also entirely avoided mass transfer under the isolated binary formation scenario \citep{Iorio2024, ElBadry2024}. In contrast, the present-day orbital separations of Gaia BH1 and Gaia BH2 appear too small to completely avoid binary interactions once the BH progenitor leaves the main sequence, at least within the framework of classical population synthesis models \citep{Kotko2024, Nagarajan2025a, Chawla2025}. We therefore focus in this work on the Gaia BH1 and Gaia BH2 systems.

Some alternative explanations based on detailed binary evolution calculations suggest that the progenitors of Gaia BH1 and Gaia BH2 could have avoided a mass-transfer phase if different calibrations of stellar evolution parameters are adopted. One possibility is that convective overshooting above the core and into the radiative envelope limits the radial expansion of massive stars, yielding substantially smaller maximum radii for supergiant progenitors \citep{Gilkis2024}. Such stars could have remained within their current orbital separations without triggering Roche-lobe overflow. Another proposed explanation involves enhanced stellar wind mass loss at high metallicity, which can simultaneously reduce the stellar radius and widen the orbit, thereby preventing binary interaction \citep{Kruckow2024A&A}. Finally, since standard binary evolution models struggle to reproduce the observed system parameters, several studies have suggested a dynamical origin for these systems. For example, hierarchical triple evolution could account for the present-day configuration if an inner-binary merger occurred and the merger product subsequently underwent efficient chemically homogeneous evolution \citep{Li2024}. Studies of young open star clusters indicate that Gaia~BH1 and BH2 form more efficiently through dynamical interactions than through isolated binary evolution, suggesting that the dynamical formation scenario may be favored \citep{Rastello2023,Tanikawa2024,DiCarlo2024}.
    
In this work, we investigate whether the observed properties of Gaia BH1 and Gaia BH2 can be reproduced through non-conservative mass transfer within the framework of isolated binary evolution. In contrast to the commonly adopted isotropic re-emission model, in which mass is lost from the vicinity of the accretor, we consider an alternative scenario where most of the transferred material leaves the system carrying the specific angular momentum of the donor’s center of mass. Under this assumption, the orbit remains constant or expands during the mass-transfer phase rather than shrinking excessively, allowing the interaction to remain stable even for systems with highly unequal mass ratios.

The unique properties of the progenitors of the Gaia BH systems appear to favor a strongly non-conservative mass-transfer scenario with low angular-momentum loss. We argue that such a mass-loss geometry is plausible given the donor’s high-opacity subsurface layers, which may enhance wind-like mass loss even prior to Roche-lobe overflow, potentially preferentially in the polar direction and producing bipolar outflows, as observed around some luminous blue variables (LBVs; \citealt{Weis2020}). Alternatively, or in addition, the strongly unequal component masses with the accompanying consequences may further contribute to this mass-loss geometry.

\begin{table}[ht] 
\centering
\caption{The parameters of Gaia BH1 and BH2: (from top to bottom) the orbital period $P $
in days, the eccentricity, BH mass $M_{\rm BH}$ and companion star mass $M_\star$ [${M}_\odot$] 
in solar units adopted from
\citet{ElBadry2023a, Chakrabarti2023, ElBadry2023b}.}
\label{tab: Gaia_parameters}
\begin{tabular}{ccc}
\toprule
Parameter & Gaia BH1 & Gaia BH2 \\
\hline
$P $ [d]              & $185.52 \pm 0.08$   & $1276.7 \pm 0.6$  \\
$e$                           & $0.439^{+0.004}_{-0.003}$ & $0.5176 \pm 0.0009$ \\
$M_{\rm BH}$ [${M}_\odot$]      & $9.32^{+0.22}_{-0.21}$    & $8.94 \pm 0.34$     \\
$M_\star$ [${M}_\odot$]         & $0.93 \pm 0.05$           & $1.07 \pm 0.19$     \\
%Companion type               & MS                        & RG                  \\
%Age [Gyr]                    & $>4.0$                    & $8.0$--$12.0$       \\
%$v_{\rm pec}$ [km\,s$^{-1}$]  & $71.3^{+9.2}_{-10.8}$      & $34.1^{+5.0}_{-5.1}$ \\
\hline
\end{tabular}
\end{table}

\section{Method} \label{sec: method}

We use the stellar evolution code {\tt MESA} \citep[version r23.05.1;][]{Paxton2011,Paxton2013,Paxton2015,Paxton2018,Paxton2019,Jermyn2023} to study the binary evolution of the progenitors of the Gaia BH1 and Gaia BH2 systems, with their parameters presented in Table \ref{tab: Gaia_parameters}. We adopt a metallicity of $Z=0.01$, which is in good agreement with both observed systems \citep{ElBadry2023a,ElBadry2023b}. The overshooting parameters for the donor star have been calibrated to match observational constraints for massive stars \citep{Brott2011}. Exponential diffusive overshooting is applied at convective boundaries: for convective cores at the top boundary ($f = 0.0425$, $f_0 = 0.001$) and for convective shells at the bottom boundary ($f = 0.02$, $f_0 = 0.004$). Convection is modeled using mixing-length theory with $\alpha_{\rm MLT} = 1.8$ and the Ledoux stability criterion. Semiconvective mixing is included with efficiency parameter $\alpha_{\rm sc} = 1.0$. 

We evolve both stars from the zero-age main sequence, including tidal interactions following \citet{Hut1980}. We assume that each stellar layer is synchronized independently, the donor has a radiative envelope, the accretor has a convective envelope, and standard tidal efficiency factors are set to unity. We present the results for the ``Kolb'' mass transfer scheme \citep{Kolb1990}. %\footnote{Note that the “roche-lobe” scheme, in which the donor’s radius is kept within its Roche lobe, yields very similar outcomes.} 
Motivated by low companion masses, we assume fully non-conservative mass transfer. In our default model, most of the mass (95\%) lost during the Roche-lobe overflow carries away the specific angular momentum of the donor's center of mass ($\alpha$=0.95, $\beta$=0.05, $\delta$=$\gamma$=0 according to the notation in \citealt{Soberman1997}). 
The angular momentum loss associated with mass lost $\dot{J}_{\mathrm{ml}}$ from the vicinity of the donor is expressed by the formula:
\[
\dot{J}_{\mathrm{ml}} = \dot{M}\, \left( a\,M_{\rm acc} / (M_{\rm acc}+M_{\rm don}) \right)^{2}\, (2\pi / P)
\]

where \(\dot{M}\) is the mass loss rate from the donor. Here, \(M_{\rm acc}\) and \(M_{\rm don}\) denote the masses of the accretor and the donor, \(a\) is the orbital separation, \(P\) is the orbital period. The factor  $ a\,M_{\rm acc} / (M_{\rm acc}+M_{\rm don}) $
 corresponds to the distance from the system's center of mass to the center of mass of the donor star.

This assumption corresponds to a spherically symmetric Jeans-mode mass loss carrying the specific angular momentum of the donor’s center of mass \citep[e.g.,][]{Huang1963,Soberman1997,Tauris2006}. In practice, this represents a spherically symmetric outflow from the surface of a non-rotating donor. Such an assumption is standard when stars lose mass through stellar winds, but it has also occasionally been applied to systems undergoing Roche-lobe overflow \citep{TutukovYungelson1971, Hurley2002}. Under this prescription, the associated angular momentum loss is relatively modest for the considered mass ratio ($M_{\rm don} \gg M_{\rm acc}$), and the orbit generally tends to expand in response to mass loss, similar to the case of isotropic stellar winds.

By contrast, if mass is lost from the vicinity of the accretor or through the outer Lagrangian points (L\(2\) or L\(3\)), in the evaluated mass ratio regime the escaping material carries away substantially larger specific orbital angular momentum \citep[e.g.,][]{Tauris2006}, leading to rapid orbital shrinkage rather than expansion.

\section{Binary Evolution} \label{sec: evolution}

\begin{figure} [ht]
    \centering
    \includegraphics[width=0.98\linewidth]{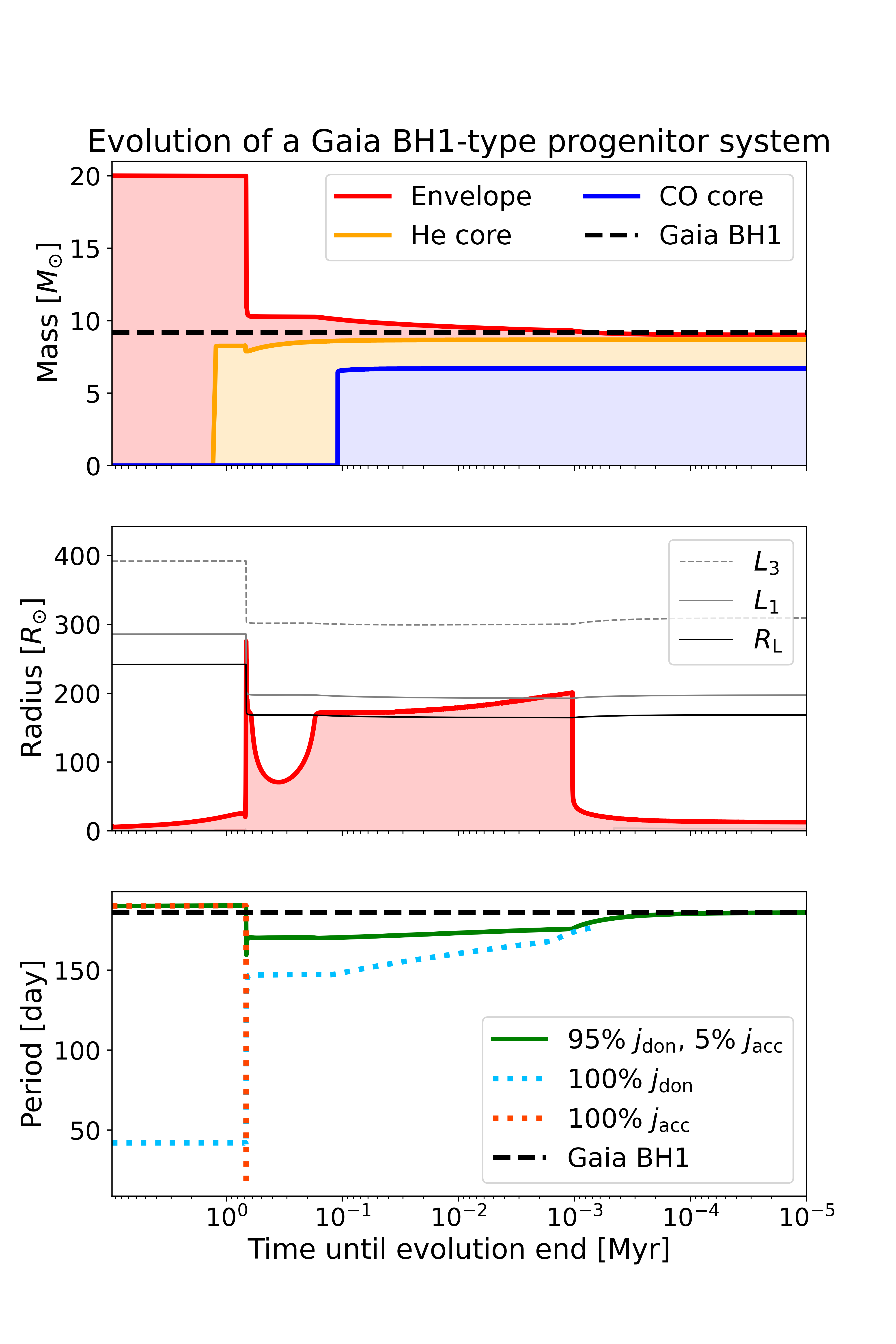}
    \caption{
    Evolution of the Gaia BH1 progenitor system. The top panel shows the evolution of the donor’s mass, with the red, yellow, and blue curves representing the hydrogen-rich envelope, helium core, and carbon--oxygen core, respectively. The middle panel presents the evolution of the donor’s radius, along with the Roche lobe radius $R_{\rm L}$ and the sizes corresponding to the inner Lagrange point L1 $\approx a \,(1 - (M_{\rm acc} / (3 M_{\rm don}))^{1/3})$ and outer Lagrange point L3 $\approx a (1 + 5/12 \, (M_{\rm acc} / M_{\rm don}))$ located behind the donor (see Fig. \ref{Fig: cartoon}). The bottom panel shows the evolution of the orbital period. The green line corresponds to the standard model, which assumes that the mass lost from the system carries away in 95\% the specific angular momentum of the donor’s center of mass ($j_{\rm don}$) and in 5\% that of the accretor’s center of mass ($j_{acc}$). For comparison, we also show two additional cases: (i) a model in which $j_{\rm acc} = 100\%$ - all of the lost mass carries the specific angular momentum of the accretor leading to rapid orbital shrinkage (orange dotted line); and (ii) a model in which $j_{\rm don} = 100\%$ - all of the lost mass carries the specific angular momentum of the donor, resulting in orbital widening (blue dotted line).} The black dashed line marks the observed properties of the Gaia BH1 system. The timescale is expressed as the logarithm of the remaining time until the end of evolution, defined as the point of central helium depletion.
    \label{Fig: Mass_evolution_1}
\end{figure} 

Figure \ref{Fig: Mass_evolution_1} presents an example of an evolutionary track leading to a system with properties similar to Gaia BH1. The initial binary system consists of a $20\,M_\odot$ BH progenitor and a $1\,M_\odot$ companion on a circular orbit with a period of 190 days, corresponding to a separation of $\sim 384\,R_\odot$. Note that the low mass companion could still be on its pre-main sequence track at the moment of primary (the more massive star) zero-age-main sequence \citep{Kippenhahn2013, Baraffe2015}. After approximately $9$ Myr, the primary evolves off its main sequence and undergoes significant radial expansion, triggering a rapid mass transfer episode lasting $\sim 0.1$ Myr with a peak rate of $3.4 \times 10^{-3}\,M_\odot\,\mathrm{yr}^{-1}$. However, as shown in the middle panel of Figure~\ref{Fig: Mass_evolution_1}, the inner Lagrange point L1 is only slightly overfilled. The outer Lagrange point L3, situated behind the donor, remains unfilled, as expected from the system geometry characteristic of the considered mass ratios \citep{Lubow1975, Eggleton1983, Ryu2025}. 

During the phase, donor rapidly loses almost $10\,M_\odot$, removing most of its hydrogen-rich envelope. Following this, the donor’s radius contracts and for over $\sim 0.4$ Myr  (most of the remaining core-He burning) the star detaches from its Roche-lobe, halting mass transfer. During this stage, the (partially) stripped donor may already appear as a WR star (see Discussion \ref{sec: discussion}). At final stages of He burning, the star expands once more, initiating a second mass transfer phase lasting $\sim 0.2$ Myr. This phase proceeds at a significantly lower rate than during the former phase, at $\sim 10^{-6}$--$10^{-4}\,M_\odot\,\mathrm{yr}^{-1}$, and results in the loss of an additional $\sim 1\,M_\odot$. Ultimately, the mass transfer ceases, leaving behind a $\sim 9\,M_\odot$ stripped helium core and a low-mass companion in a widened orbit with a period of around 170 days. Subsequently, the stripped star may undergo core collapse to a BH, accompanied by low, asymmetric mass loss, for example through neutrino stream emission \citep{Fryer1999}. This can lead to a modest orbital widening and induce an eccentricity in the binary system; see broader discussion on the possible origin of eccentricity in Section~\ref{sec: eccentricity} of the Appendix.

Within our default model, we assume that the mass lost from the system carries away 95\% of the specific angular momentum of the donor’s center of mass and 5\% of that of the accretor’s center of mass (green line in Fig.~\ref{Fig: Mass_evolution_1}). Under this assumption, the orbital separation remains nearly constant throughout the evolution. However, the orbital evolution is highly sensitive to the adopted prescription for angular momentum loss. For comparison in Fig. ~\ref{Fig: Mass_evolution_1} we consider two additional cases: (i) a model in which 100\% of the lost mass carries the specific angular momentum of the accretor, leading to rapid orbital shrinkage and likely stellar merger or common envelope phase (orange dotted line), and (ii) a model in which 100\% of the lost mass carries the specific angular momentum of the donor, resulting in constant orbital widening (blue dotted line). In particular, if the fraction of mass lost with the higher specific angular momentum characteristic of the accretor’s vicinity exceeds $\sim 6\%$, the mass transfer becomes unstable. The evolutionary outcome is also affected by the assumptions adopted for tidal interactions.

The evolutionary pathway obtained for the Gaia BH2 system is qualitatively similar to the scenario presented for Gaia BH1, differing primarily in its much wider orbital separation. In the case of the Gaia~BH2 system, the donor star is therefore significantly more expanded at the moment it fills its Roche lobe, and after the system becomes detached it does not fill the Roche lobe again. A detailed, analogous description of the Gaia~BH2 progenitor system is provided in Appendix~\ref{Appendix}.

{\section{Physical Motivation for the Model} \label{sec: mechanism}}

\begin{figure*} [ht]
    \centering
    \includegraphics[width=1.0\linewidth]{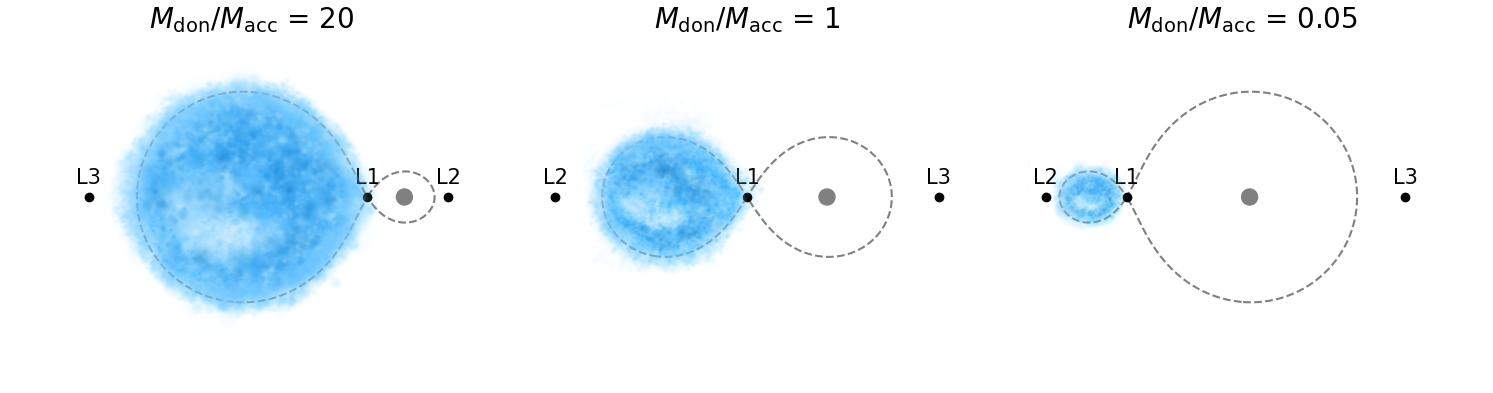}
    \caption{Illustration of the geometry of systems at the onset of Roche-lobe overflow for three different mass ratios, all with the same orbital separation and total components mass. The left panel shows a system with a donor-to-accretor mass ratio $M_{\rm don}/M_{\rm ac}=20$, as in case of Gaia BH1 and BH2 progenitors. For comparison, cases with equal-mass components ($M_{\rm don}/M_{\rm ac}=1$) and with the opposite extreme mass ratio ($M_{\rm don}/M_{\rm ac} =$0.05) are also presented.
    }
    \label{Fig: cartoon}
\end{figure*} 

%\begin{figure} [ht]
 %   \centering
  %  \includegraphics[width=0.97\linewidth]{Accretor_surface.png}
   % \caption{Scaling of the accretor’s effective Roche-lobe cross-section with mass ratio $q = M_{\mathrm{don}}/M_{\mathrm{acc}}$. The plotted quantity is $\sigma_{\mathrm{acc}}(q)/\sigma_{\mathrm{acc}}(q=1) = [f(1/q)/f(q)]^2$, where $f(q)$ is the Roche-lobe approximation \citep{Eggleton1983}. Crosses indicate representative mass ratios at the beginning $q=20$ and the end $q\approx 9$ of evolution. }
    %\label{Fig: accretor_surface}
%\end{figure} 

There are several reasons to expect enhanced mass loss from the vicinity of the donor in wide, highly unequal-mass systems with a massive donor ($M_{\rm don}> 20 M_{\odot}$), such as those studied in this work. Below, we outline some factors that, separately or combined, could lead to the assumed mass-loss geometry, thereby limiting angular momentum loss and preventing unstable mass transfer in the progenitors of the Gaia BH systems.

\subsection{Unequal mass ratio}

First, the mass ratio of the binary does affect the gas dynamics of Roche-lobe overflow through its influence on the Roche potential near the L1 point, which determines the geometry of the equipotential surfaces and the properties of the mass transfer stream \citep{Lubow1975}. On top of this, we introduce our simple geometrical argument in the following.

Given the highly unequal mass ratios in these systems, a fraction of the mass-transfer stream launched near the inner Lagrange point L1 may not be confined within the accretor’s Roche lobe due to its small size. Figure~\ref{Fig: cartoon} illustrates the geometry of binaries at the onset of Roche-lobe overflow for three different mass ratios, all assuming the same orbital separation and total system mass. The left panel shows a system with a donor-to-accretor mass ratio of $M_{\rm don}/M_{\rm acc}=20$, representative of the progenitors of Gaia BH1 and BH2. In such highly unequal-mass-ratio binaries, the accretor’s Roche lobe is very small compared to the donor’s surface.

The geometric cross-section of the accretor’s Roche lobe as seen by the incoming mass stream changes significantly with mass ratio. Approximating the Roche-lobe radius using the formula by \citet{Eggleton1983} $R_{\rm L, don} = f(q)\,a,$ the accretor's Roche-lobe radius can be expressed as $R_{\rm L,acc} = f(1/q)\,a$. Thus the ratio of accretor cross-sections between unequal and equal-mass systems scales as $\left[{f(1/q)}/{f(q)}\right]^2$.
%As illustrated in Figure~\ref{Fig: accretor_surface}, 
At the beginning of the Roche-lobe overflow in our systems with $q=20$ the accretor presents only $\sim 7\%$ of the cross-section it would have in an equal-mass binary. If we assume that the accretion efficiency $\beta$ is roughly proportional to this effective cross-section - for example, if the mass stream spreads over a finite solid angle depending on its perpendicular velocity at L1 - then a system that would transfer mass nearly conservatively at $q \approx 1$ ($\beta \sim 50$--$100\%$) would exhibit $\beta \sim 3$--$7\%$ at $q \sim 20$. 

This simple geometric argument provides an intuitive motivation for adopting low accretion efficiencies in highly unequal-mass systems.
We speculate that during the mass-transfer phase, a fraction of the material is transferred through L1. However, once the accretor’s Roche lobe becomes filled, or if the mass-loss rate becomes sufficiently high, this channel may be effectively blocked, limiting the amount of material that can enter the accretor’s Roche lobe and increasing the importance of alternative mass-loss channels.

\subsection{Self-accretion}
Moreover, in systems with highly unequal masses ($M_{\rm acc} \ll M_{\rm don}$), such as our Gaia BH progenitors, low initial stream velocities and asynchronous donor rotation can significantly modify the mass-loss geometry, potentially leading to substantial self-accretion \citep{davisMassTransferEccentric2013,davisBinaryEvolutionUsing2014,Hendriks2023}. In particular, if the donor rotates sufficiently more slowly than the orbit—i.e., with an angular velocity below approximately 50\% of the orbital rate \citep{Hendriks2023}—the L1 stream may entirely miss the accretor’s Roche lobe and instead fall back onto the donor. This process may produce a non-standard mass-loss geometry. Self-accretion can lead to enhanced accumulation of material near the Roche-lobe boundary, potentially increasing mass loss from the donor’s outer layers and, in certain regions of parameter space (depending on mass ratio and synchronization), even leading to orbital widening. In luminous massive stars, radiative forces may further influence the dynamics of this material.

However, the exact impact of self-accretion on the angular momentum budget of the system—and consequently on orbital evolution and donor spin evolution—remains unexplored. It is unclear whether such effects could mitigate the strong orbital shrinkage predicted by standard models. These processes are not accounted for in the simplified mass-transfer prescriptions commonly adopted in 1D binary evolution models, although they may significantly affect both the orbital evolution and the mass-growth history of binary systems.

\begin{figure*} [ht]
    \centering
    \includegraphics[width=0.45\linewidth]{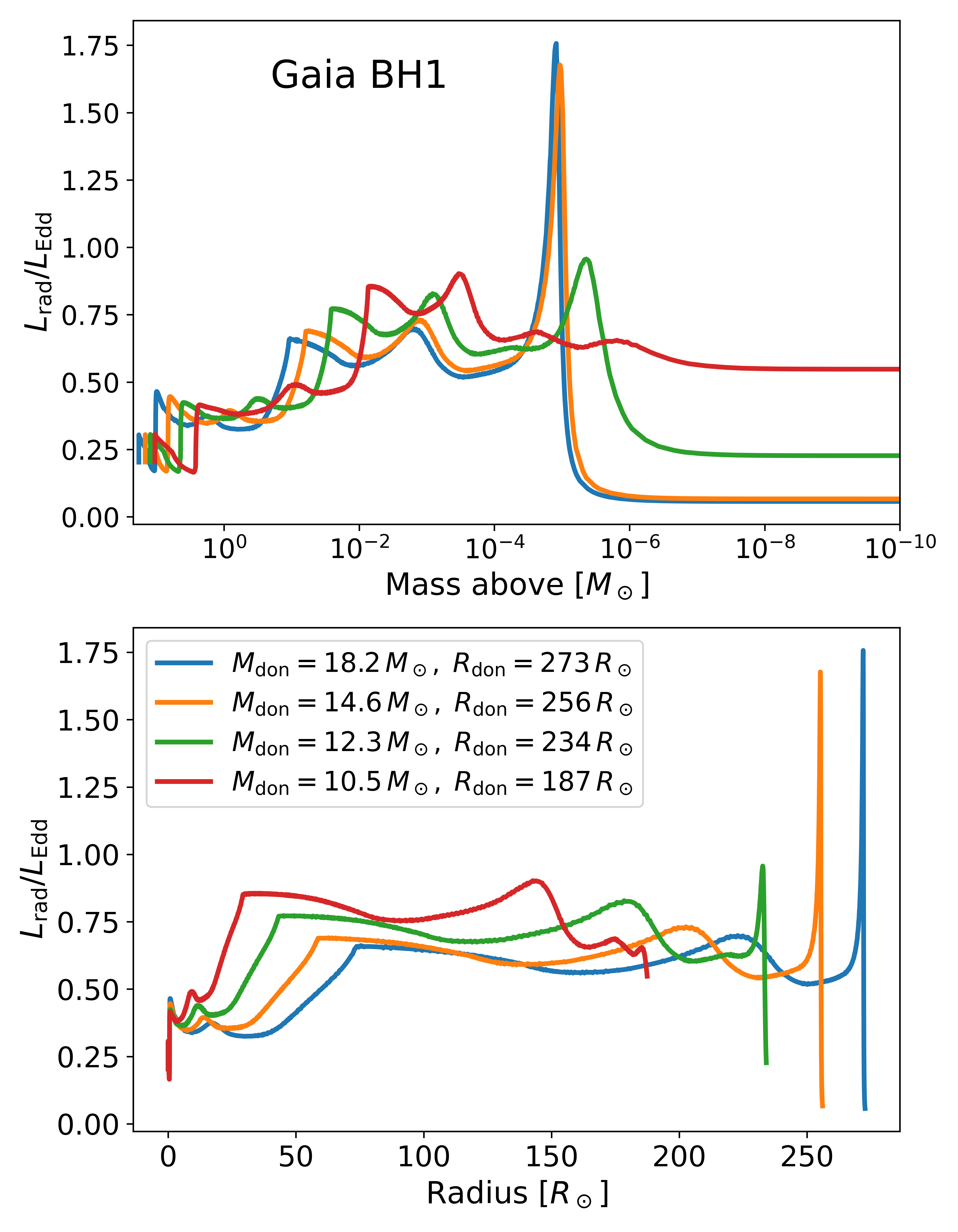}
    \includegraphics[width=0.45\linewidth]{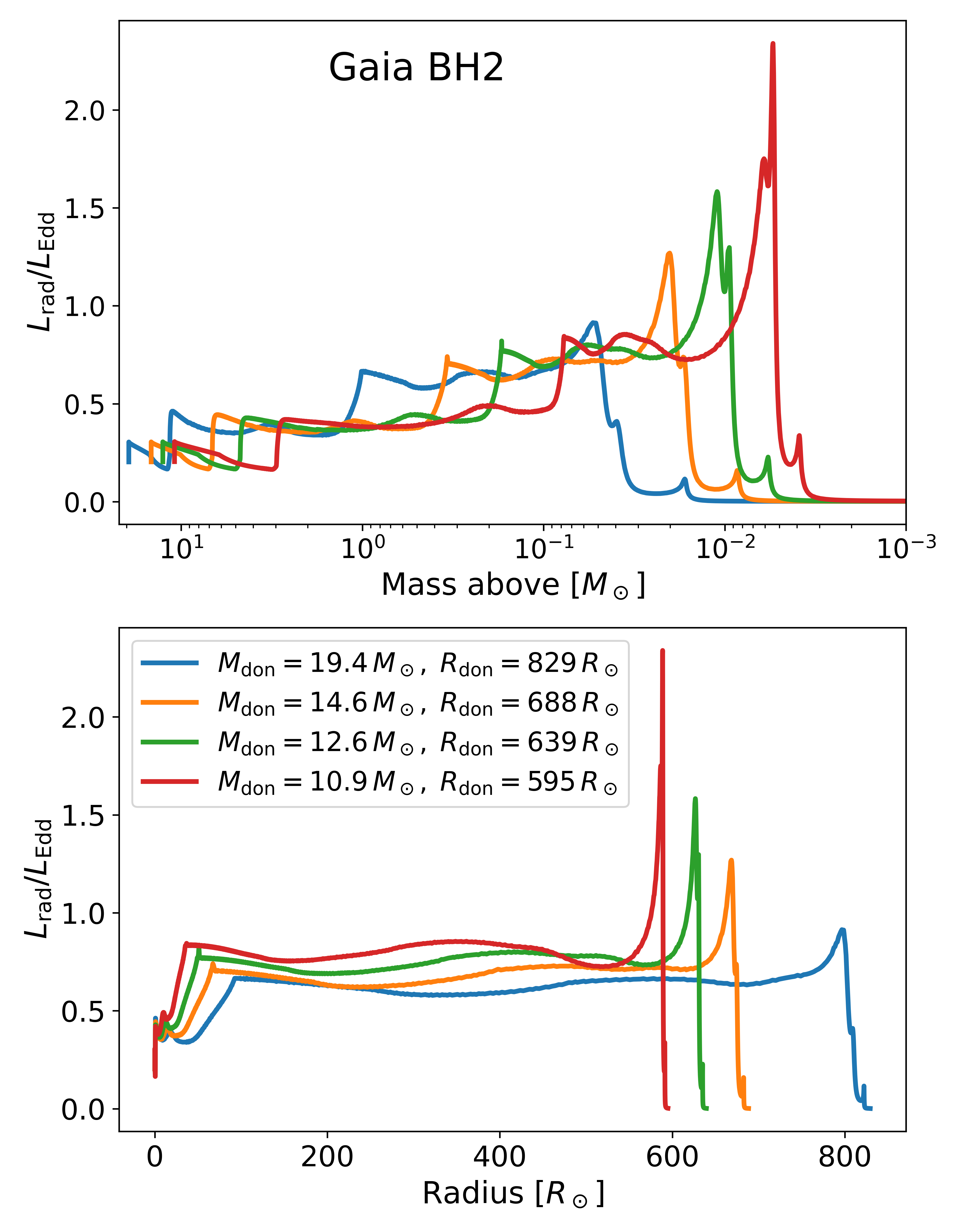}
    \caption{Luminosity profiles of donor stars - Gaia BH1 (left) and BH2 (right) progenitors from {\tt MESA} simulations, shown as a function of mass above a given layer (top panels) and stellar radius (bottom panels). Results are shown for four selected evolutionary stages during ongoing mass loss, with each curve labeled by the donor star's total mass $M_{\rm don}$ and radius $R_{\rm don}$. Luminosity in each profile is expressed as the ratio between the local radiative luminosity 
    and the Eddington luminosity, $L_{\rm rad}/L_{\rm Edd}$, where super-Eddington layers 
    are those with $L_{\rm rad}/L_{\rm Edd} > 1$. For rotating stars, these limits would be lower due to the centrifugal reduction of the effective gravity \citep[e.g.,][]{MaederMeynet2000}.
 .}
    \label{Fig: Luminosity}
\end{figure*} 

\subsection{High-opacity layers and bipolar mass loss}

The ideal scenario for the angular momentum budget assumed in our models would be if the majority of the mass was lost from the donor in the form of an enhanced wind.
Massive stars, particularly at moderate to high metallicities, develop high-opacity subsurface layers in which the local luminosity can exceed the Eddington limit \citep{Joss1973,Jiang2015,Sanyal2015,Sanyal2017,Jermyn2023}. Such locally super-Eddington conditions may unbind the overlying material and drive strong, eruptive mass loss at rates of up to $\sim 10^{-2} \, M_{\odot} \, {\rm yr}^{-1}$\citep{Owocki2015,Quataert2016,Cheng2024,Pauli2026}, potentially explaining the observed paucity of red supergiants above $\gtrsim 30 \,M_{\odot}$. The amount of mass in sub-surface super-Eddington layers depends on the effective temperature of the star \citep{Cheng2024} and is the largest for cool donors in wide orbits. 

Interestingly, similar high opacity peaks appear in the stellar structure profiles of our Gaia BH1 and BH2 progenitors at the onset and during the ongoing mass transfer phase, suggesting that analogous processes could have contributed to their formation. 
Figure~\ref{Fig: Luminosity} shows the luminosity profiles of donor stars 
from {\tt MESA} simulations for the Gaia BH1 (left) and Gaia BH2 (right) progenitors, 
plotted as a function of mass above a given layer (top panels) and stellar radius (bottom panels). Results are shown for four selected evolutionary stages during ongoing mass loss, 
with each curve labeled by the donor star's total mass $M_{\rm don}$ and radius $R_{\rm don}$. In both cases near-surface layers reach luminosities exceeding the Eddington limit, implying that these layers are formally unbound even for a non-rotating single star 
due to strong radiative pressure. If the donor is rotating (centrifugal support) or in the presence of a companion, the corresponding Eddington limit is even lower as effective gravity is reduced \citep{MaederMeynet2000}. The opacity peaks shown in Fig.~\ref{Fig: Luminosity} give rise to subsurface convective zones \citep{Cantiello2009}, in which tidal dissipation is expected to be much stronger than in the rest of the radiative envelope \citep{Zahn1977,Ogilvie2014}. This could lead to these layers becoming fully or partly synchronized with the orbit by tides, thereby further reducing their effective gravity. 

Figure~\ref{Fig: Luminosity} shows that the high-opacity subsurface regions with $L_{\rm rad} > L_{\rm Edd}$ persist in our donor models throughout nearly the entire mass-loss phase. This behavior is not unexpected, as the thermal timescale of these layers is sufficiently short to allow the envelope structure—including the partial-ionization zones—to relax quasi-continuously. Similar effects have previously been discussed in the context of mass-transfer stability and the subsurface entropy structure of mass-losing stars \citep{Pavlovskii2015,Temmink2023,Klencki2025}. For the more extended Gaia~BH2 progenitor, the subsurface layers exceed the Eddington limit even more strongly as the star becomes increasingly stripped. In addition, the high-opacity regions are located deeper in the Gaia~BH2 progenitor ($\sim 5$--$25~R_\odot$) than in the Gaia~BH1 progenitor ($\sim 1~R_\odot$). 

Recently, \citet{Cehula2025} demonstrated that radiation pressure can substantially enhance mass loss from massive donors in Roche-lobe-underfilling binaries. In particular, donors with super-Eddington, convectively inefficient subsurface layers can drive mass transfer at rates of $\geq 10^{-2} M_{\odot}$yr$^{-1}$ even before they overflow their Roche lobes.
While some of the excess luminosity in layers with $L_{\rm rad}/L_{\rm Edd} > 1$ may be transported by convection rather than driving outflows, 3D radiation--hydrodynamic simulations of stellar envelopes show that convection becomes inefficient near the stellar surface, where the photon diffusion timescale becomes shorter than the dynamical (convective) timescale \citep{Jiang2015,Goldberg2022,Ma2025}. 

Note that our donor stars could represent LBVs. Indeed, several LBVs are observed in wide binary systems with eccentric orbits, and their enhanced mass loss is often attributed to various forms of binary interaction, depending on the system (see, e.g., \citealt{Smith2026} and references therein). A significant fraction of LBVs ($\sim$60--70\%) exhibit varying degrees of bipolarity, from pronounced hourglass-shaped nebulae to weaker bipolar caps \citep{Smith2006,Weis2011,Weis2020}, likely tracing past episodes of polar-enhanced mass loss. Such bipolar outflows provide a favorable geometry for mass loss in the Gaia BH progenitor systems considered here, as they preferentially remove matter with low angular momentum, thereby helping to prevent unstable mass transfer.

Bipolar mass loss scenario and observations of LBVs are further supported by theoretical considerations, particularly the effects of gravity darkening described by the von Zeipel theorem \citep{vonZeipel1924a,vonZeipel11924b}. In rotating stars, the effective gravity at the equator is reduced due to rotation \citep{MaederMeynet2000} and potentially enhanced by tidal interactions with a close companion. Since the local radiative flux scales with the effective gravity, mass loss may be stronger at the poles than near the equator, producing bipolar outflows \citep{Owocki1997,MaederMeynet2000}. 

Taken together, observational and theoretical evidence suggests that bipolar outflows may play an important role in the evolution of systems with massive stars and influence the formation of compact-object binaries. However, verifying the occurrence of this mechanism in some binaries and assessing its detailed impact on their evolution would likely require multidimensional stellar models, which remain beyond the current state of the art.

\subsection{Other}

If the system were initially eccentric, it could retain part of this eccentricity throughout the mass-loss phase, as suggested by observations of wide binaries \citep{Oomen2018, Oomen2020, Escorza2020, Muller-Horn2025}. The evolution of systems undergoing eccentric mass transfer remains poorly understood, but recent theoretical modelling \citep{Rocha2025, Parkosidis2025, Parkosidis2025b} indicates that it may result in more orbital widening than predicted for circular systems. In the mass-ratio regime of our systems, stable mass transfer nevertheless requires substantially lower angular momentum loss—such as specific angular momentum of the donor’s center of mass—compared to standard binary evolution prescriptions (i.e., isotropic re-emission).

It is worth noting that in the past some studies considered mass-loss geometries in which the lost material carries away the specific angular momentum of the donor’s center of mass during Roche-lobe overflow \citep[e.g.,][]{TutukovYungelson1971, Vanbeveren1991, Vanbeveren1998}. This treatment was adopted, for example, as the default prescription in the original formulation of \citet{Hurley2002}.

\section{Discussion} \label{sec: discussion}

In this Letter, we present a new evolutionary model that successfully reproduces the formation of the Gaia BH1 and BH2 systems. Our scenario assumes fully non-conservative mass transfer, with most of the material ($\gtrsim 95-100 \%$) lost from the system carrying the specific angular momentum of the donor’s center of mass. We argue that such a mass loss mode from the vicinity of donor is physically plausible in wide, unequal-mass binaries with massive donors like the Gaia BH progenitors due to several independent factors: possible eruptive mass loss driven by high-opacity subsurface layers, unequal Roche-lobe sizes or sub-synchronous donor rotation, and self-accretion (see Sec.~\ref{sec: mechanism}). At present, it is unclear which mechanism, or combination of mechanisms, provides the primary source of low–angular-momentum mass loss. We note also that while scenario with bipolar outflows from high-opacity layers favor rotating donors, self-accretion requires sub-synchronous rotation. Determining the dominant process is beyond the state of the art and capabilities of current 1D stellar-evolution models. Although our prescription is idealized and may not capture well the exact details, such as e.g., mass-loss rate, it provides a motivation for developing more complex mass-transfer models that account for the properties of the progenitors. In Section \ref{sec: caveats}, we outline possible caveats of the adopted methodology and discuss the limitations of the proposed scenario. 

In Section \ref{sec: BH_population}, using population synthesis, we estimate the formation efficiency of Gaia BH1- and Gaia BH2-like systems and discuss the implications of the adopted angular momentum loss prescriptions, comparing them with a standard model. In Section \ref{sec: broader_implications}, we discuss the possible impact on other post-mass-transfer binaries and outline future directions.

\subsection{Caveats} \label{sec: caveats}

In the presented evolutionary scenario, we assume that both components of the binary form simultaneously. In reality, however, the pre-main-sequence evolution of the low-mass companion may play an important role. For instance, the time required for a $1 {M}_\odot$ star to reach the zero-age main sequence can be delayed by several megayears, which is comparable to the lifetime of its massive companion. During this phase, the stellar radius may be several times larger than on the main sequence \citep[e.g.,][]{Kippenhahn2013, Baraffe2015}, potentially leading to an episode of early mass transfer. Nevertheless, the formation of such systems at these early stages is likely more complex and may involve additional factors and processes, such as dynamical interactions.

Most popular prescriptions for non-conservative mass transfer in binary evolution assume that the lost mass carries the specific angular momentum of the accretor’s center of mass. In this work, we instead adopt a model in which the ejected mass carries the specific angular momentum of the donor’s center of mass. In reality, mass loss can occur through more complex geometries. Recent studies suggest that, under certain rapid mass-transfer conditions, some material may also escape through the L3 (behind the donor) or L2 (behind the accretor) Lagrange points \citep{Lu2023, Scherbak2025}. Such mass loss would carry substantially more angular momentum than that associated with the donor’s center of mass. For the mass ratios considered here, where the donor is significantly more massive than the accretor, an L3 outflow is unlikely \citep{Lubow1975, Eggleton1983, Ryu2025}. Indeed, as illustrated in Figures~\ref{Fig: Mass_evolution_1} and~\ref{Fig: Mass_evolution_2}, the L3 point remains unfilled throughout our simulations. A very small fraction of the total mass being lost through the outer Lagrangian point does not necessarily lead to instability \citep{Marchant2021}, but such cases should nonetheless be treated with caution.

Note that \citep{Kruckow2024A&A} also proposed mass loss with low specific angular momentum via strong stellar winds as a possible formation pathway for Gaia BH1 and BH2; however, their scenario differs in important respects from the one presented here. In \cite{Kruckow2024A&A}, the authors searched for evolutionary solutions in which the orbit widens significantly early in the evolution, thereby preventing the donor from filling its Roche lobe at all. This requirement led to very high constraints for the initial BH progenitor masses of $\geq 80\,M_{\odot}$, necessary to produce sufficiently strong wind-driven mass loss before the star expands too much.
In contrast, we consider enhanced mass loss to occur shortly before and during the Roche-lobe overflow. At this stage, the donor stars are substantially more expanded, which allowed them to develop high-opacity, super-Eddington subsurface layers potentially capable of driving eruptive mass loss. In our standard model, we also allow a fraction of the transferred material ($\sim 5\%$) to reach the vicinity of the accretor and then be lost with much higher specific angular momentum than in the wind case. This leads to a more complex orbital evolution governed by the interplay between tides and mass loss, sometimes resulting mild orbital shrinking or widening. Overall, however, in our default model the orbital separation remains approximately constant throughout evolution of the system. Importantly, our scenario allows for much lower BH progenitor masses ($\sim 20$--$30\,M_{\odot}$) than in \cite{Kruckow2024A&A}.

All Gaia BH systems detected to date are eccentric, which may have influenced their mass-transfer evolution, depending on whether the eccentricity was inherited from earlier evolutionary stages or induced after BH formation. Possible origins of the eccentricity, as well as the potential to constrain factors such as natal kicks, are discussed in more detail in Appendix~\ref{sec: eccentricity}.

Another potential caveat is that the progenitors of Gaia BH systems are often thought to lie within the regime of the so-called Darwin instability, in which tidal synchronization extracts angular momentum from the orbit and can potentially lead to a stellar merger \citep{Darwin1879, Zahn1977, Hut1980}. Our {\tt MESA} models include tidal interactions, and under our standard assumptions, the orbit expands efficiently in response to mass transfer, avoiding the Darwin instability. The ultimate fate of the system and the possibility of a merger, however, depend on the tidal prescription and the adopted tidal efficiency, which remain highly unconstrained.

\subsection{Effects of angular momentum loss on Gaia BH1 and BH2 formation efficiency} \label{sec: BH_population}

\begin{figure} [ht]
    \centering
    \includegraphics[width=0.98\linewidth]{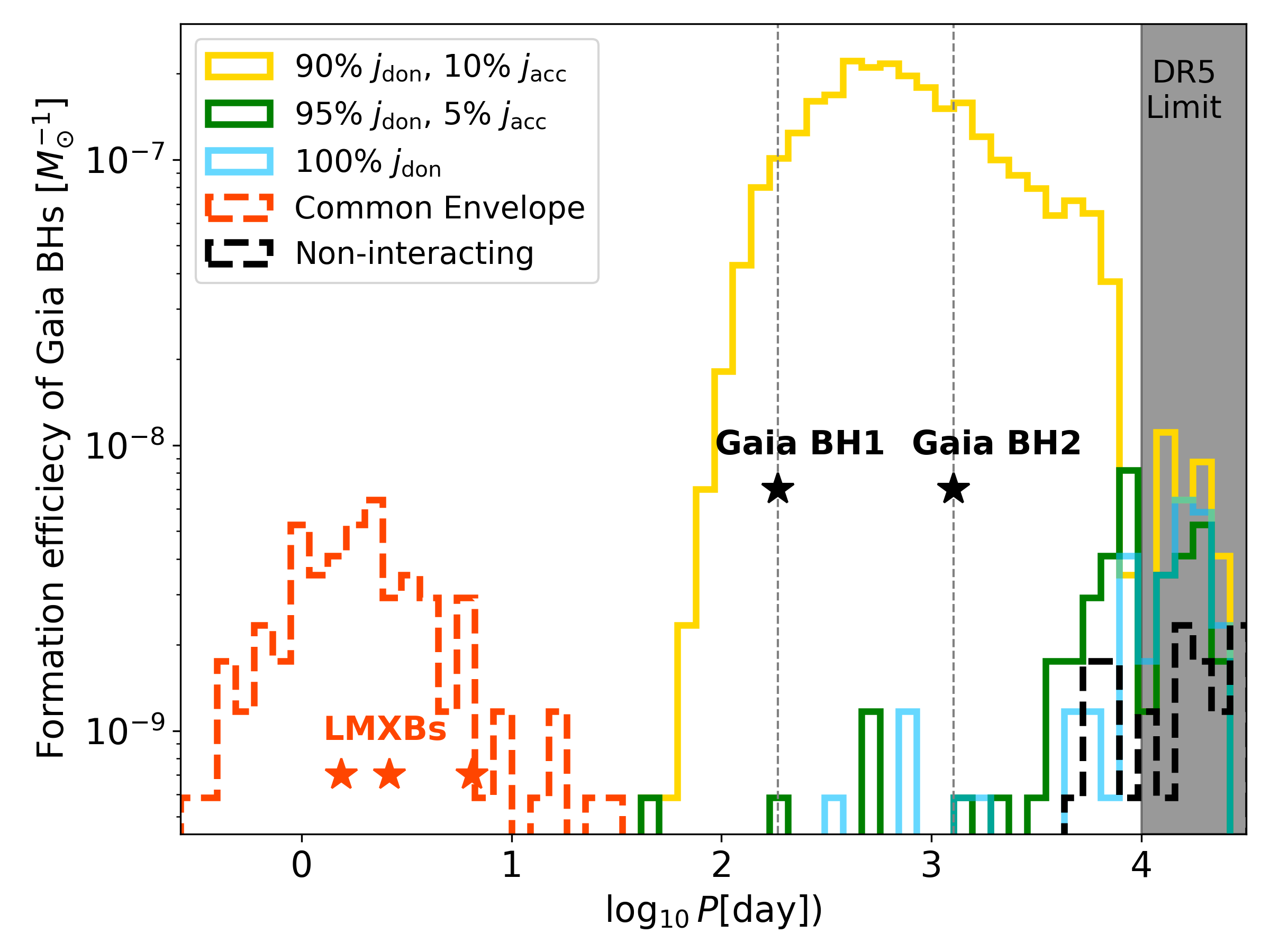}
    \caption{Formation efficiency of Gaia BH-like systems per solar mass unit at metallicity $Z=0.01$ as a function of an orbital period. Gaia BH-like systems are defined as binaries containing a BH with mass $8$--$10,M_\odot$ and a companion star with mass $0.5$--$1.5\,M_\odot$ (see Sec.~\ref{sec: BH_population}). Results of \texttt{StarTrack} population synthesis simulations with different assumptions regarding angular momentum loss during non-conservative mass transfer. Solid lines correspond to models with low-angular-momentum mass loss, in which the ejected material carries the donor’s specific angular momentum at levels of 90\% (yellow), 95\% (green), and 100\% (blue). For comparison, we also include the standard \texttt{StarTrack} model (orange dashed line), in which all progenitor systems undergo unstable mass transfer treated with the energy-balance common-envelope prescription with $\alpha_{\rm CE}=1$. Black markers indicate the observed Gaia BH1 and Gaia BH2 systems. Orange markers show the orbital periods of three low-mass X-ray binaries (LMXBs): V404 Cygni \citep{Khargharia2010}, GRO J1655--40 \citep{Shahbaz1999}, and XTE J1550--564 \citep{Orosz2011}.}
    \label{Fig: Period_distributions}
\end{figure} 

In this section, we explore the effect of the assumed mass-loss geometry on the formation efficiency of Gaia BH1- and BH2-like systems, defined as binaries that reproduce the observed masses and orbital separations of the two systems. We use \texttt{StarTrack} population synthesis code \citep{Belczynski2008,Belczynski2020} and test different variants of the angular momentum loss during fully non-conservative mass transfer. Three of them correspond to the low-angular-momentum mass loss geometry in which the ejected material carries the donor’s specific angular momentum $j_{\rm don}$ in 90\%, 95\%, and 100\% fractions. The remaining mass-loss (if any) carries away the accretor’s specific angular momentum. 

For comparison, we also include the standard mass transfer model in which all progenitor systems, due to their extreme mass ratio, undergo an unstable mass transfer phase, even within the revised stability criteria \citep{Olejak2021a}. For systems undergoing unstable mass transfer, we adopt an optimistic scenario in which they avoid merger and can successfully survive the common-envelope phase. Their subsequent evolution is computed using the energy-balance common-envelope prescription with $\alpha_{\rm CE}=1$ \citep{1984ApJ...277..355W}. For the BH masses, we adopt the supernova engine prescription of \citep{Fryer2022}, with $f_{\rm mix}=4.0$, corresponding to rapid convective growth.

We generate populations of binary systems with metallicity $Z=0.01$. The initial masses of the binary components are drawn from the initial mass function of \citet{Kroupa1993}. However, we evolve only $200\,000$ systems with primary masses corresponding to BH progenitors in the range $20$--$300\,M_\odot$ and secondary masses in the range $0.08$--$2.0\,M_\odot$. Initial orbital periods are sampled following \citet{Sana1207} from a power-law distribution in $P_0=\log_{10}(P_{\rm orb}/{\rm day})$, such that $f(P_0)\propto P_0^{p}$, with exponent $p=-0.55$. We restrict the period range to $P_0\in[0.05,\,4.0]$, corresponding to binaries expected to undergo binary interaction during their evolution.

From the evolved population, we select systems that form Gaia BH1- and Gaia BH2-like binaries, defined as systems hosting BHs with masses in the range $8$--$10\,M_\odot$ and low-mass companion stars with masses in the range $0.5$--$1.5\,M_\odot$. We then normalize the number of such systems to the total stellar mass represented by the underlying initial mass function, yielding the formation efficiency of Gaia BH-like binaries in units of systems per solar mass of star formation at $Z=0.01$.

The results for the tested models are presented in Figure~\ref{Fig: Period_distributions}. The model in which $90\%$ of the ejected material carries the specific angular momentum of the donor (yellow line) produces a broad peak in the orbital-period distribution in the range consistent with the observed Gaia BH1 and BH2 systems, i.e., $150$--$1500$ days. This model forms Gaia BH1- and BH2-like systems with an efficiency approximately three orders of magnitude higher than the other tested models. 

In contrast, the models in which the fraction of mass lost carrying the donor's specific angular momentum $j_{\rm don}$ is increased to $95\%$ (green line) and $100\%$ (blue line) tend to produce systematically wider binaries. The two models also yield fewer BH systems due to the increased probability of binary disruption during BH formation. Separations of the widest BH systems produced in the three models with 
$j_{\rm don}$ mass loss fractions $: 90\%$, $95\%$, and $100\%$ partly overlap with the population of non-interacting binaries that did not initiate a mass-transfer phase during their evolution (black, dashed line). These systems have orbital periods above the Gaia DR3, DR4, and the future DR5 detectable ranges, with cutoffs at approximately 4000, 7000, and 10000 days, respectively \citep{Nagarajan2025a}. Note that systems with periods significantly longer than the observing baseline are undetectable. 

The wider separations in the {\tt StarTrack} population, which differ from the {\tt MESA}-based expectations, are likely driven by two main factors. One of them is the treatment of convective core overshooting and, therefore the resulting differences in core-to-envelope mass fractions between the two codes. Consequently, lower core masses in {\tt StarTrack} and more massive envelopes lead to larger in total mass lost from the system and also wider post-interaction systems. Discrepancies also arise from differences in the tidal effects between the two codes, see Section ~\ref{sec: method} and \citep{Belczynski2020}. As shown in Section~\ref{sec: evolution}, the interplay between mass-loss-driven widening and tides in {\tt MESA} simulations determines the relatively constant orbital period obtained in the case where 95\% of the mass is lost, carrying the donor’s specific angular momentum. This subtle effect is not reproduced by the equivalent model in {\tt StarTrack}. 

As expected, the common-envelope channel (orange dashed line in Fig.~\ref{Fig: Period_distributions}) predominantly produces short-period binaries, with orbital periods ranging from a few hours up to $\sim 10$ days. This leads to a pronounced deficit of post-common-envelope systems with orbital periods comparable to those of Gaia BH1 and Gaia BH2. The Gaia detectability of short-period systems typical for this model is strongly limited by selection effects, restricting observations to relatively nearby systems \citep{ElBadry2024c}. 
Nevertheless, such BH population does exist and is observed primarily in the form of low-mass X-ray binaries, such as V404 Cygni \citep{Khargharia2010}, GRO J1655--40 \citep{Shahbaz1999}, and XTE J1550--564 \citep{Orosz2011}. These systems host BHs and donor stars with similar masses with those of Gaia BH1 and BH2. Low-mass X-ray binaries, as shown in Figure \ref{Fig: Period_distributions}, are not reproduced in the alternative mass-loss prescriptions assuming low specific angular momentum loss. This discrepancy may indicate that multiple formation channels operate in nature.

Making a rough assumption that the mass of the Galactic disk is $5 \times 10^{10}\,M_\odot$ \citep{Licquia2015}, the $90\%\,\,j_{\rm don}$ model produces $\sim 10^{5}$ Gaia BH-like systems with orbital periods between 100 and 1500 days. The other models: 90\% and 100\% angular momentum loss prescriptions, produce $\sim 10^{2}$ such systems, while the common-envelope evolution channel produces none in this period range.
Our results are consistent with recent independent studies by \cite{Xu2026} and \cite{Mapelli2026}, which investigated similar population-synthesis studies and likewise found that mass loss dominated by the donor's specific angular momentum is a promising and the most efficient mechanism for reproducing the observed Gaia BH systems via isolated binary evolution. Note that analysis in this section is restricted to BH binaries with component masses similar to those of Gaia BH1 and Gaia BH2. In contrast, the Galactic BH population is significantly more diverse, spanning a wide range of component masses, orbital configurations, and companion types \citep{Olejak2020}. Moreover, the currently observed sample is subject to strong observational selection effects, which preferentially favor the detection of specific classes of systems \citep{ElBadry2024c}. 

The pronounced sensitivity of the predicted population to relatively small variations in the low angular momentum mass-loss fraction, such as between 90\% and $95\%$ of $j_{\rm don}$, together with discrepancies between {\tt MESA} and {\tt StarTrack}, highlights the need for careful interpretation and implementation of such prescriptions in population-synthesis studies. In particular, the geometry, stability, and efficiency of mass loss may depend sensitively on the properties of the binary system (see the following subsection). Future work should therefore explore the impact of these assumptions over a broader parameter space and across different BH populations.

\subsection{Broader implications} \label{sec: broader_implications}

If the mass-transfer model proposed here represents a realistic formation scenario for the Gaia BH systems, it would have broad implications for other binary populations. While some post-mass-transfer binaries appear to undergo relatively conservative early-phase mass transfer, well reproduced by standard assumptions \citep[e.g.,][]{Lechien2025, Sen2025}, several systems exhibit puzzlingly wide orbital separations similar to those of the Gaia BH systems, challenging the popular isotropic re-emission model. A mass-loss mode carrying the donor’s specific angular momentum could help to explain some of these cases. If similar mass loss occurs (perhaps less efficiently) in more equal-mass or lower-mass binaries, it could substantially reshape the predicted properties of all post-mass-transfer populations. Here, we provide a few illustrative examples.

Similar mass-loss may contribute to the formation of relatively wide white dwarf \citep{Yamaguchi2025} and neutron star binaries \citep{ElBadry2024b} detected by Gaia. In some of these systems, the observed orbital periods, similarly to BH systems, challenge standard isolated binary evolution scenarios and the CE scenario. 

Another intriguing class of systems that appear inconsistent with predictions from standard binary evolution models are the stripped Wolf–Rayet stars observed in the Small Magellanic Cloud \citep{Schootemeijer2024}. Monitoring of seven known cases led to the surprising conclusion that none host a close, massive companion capable of removing the hydrogen envelope. Given the relatively low metallicity of the Small Magellanic Cloud, purely wind-driven mass loss would be insufficient to strip the envelope unless enhanced by eruptive episodes. We argue that, although the survey ruled out massive, close companions, these stripped Wolf–Rayet stars may still harbor low-mass companions on wide orbits. In our scenario, radial velocities of post-mass-transfer stripped helium stars are of order $\sim 4$–8~km~s$^{-1}$. \citet{Schootemeijer2024} found low radial-velocity variations between 6 and 23~km~s$^{-1}$ (median standard deviation of 5~km~s$^{-1}$) for the seven Wolf–Rayet stars, suggesting that systems with low-mass companions could remain undetected due to similarly small velocity amplitudes. We speculate that, analogous to the progenitors of Gaia BH systems, these stars may have undergone a similar episode of orbital widening during a past Roche-lobe overflow phase.
  
The presented mass loss scenario may also have important implications for the formation of low-mass X-ray binaries with BH companions, potentially helping to resolve one of the long-standing challenges in their evolutionary formation. In traditional CE channels, the BH progenitor’s envelope is typically too tightly bound to allow successful envelope ejection, preventing the survival of the observed binary systems \citep{Kalogera1999, Podsiadlowski2003, Wiktorowicz2014, Wang2016}. In the presented evolution, the donor can expand significantly during (Fig. \ref{Fig: Mass_evolution_1} and Fig. \ref{Fig: Mass_evolution_2}). For systems with initially wider orbital periods than studied here, progenitors of Gaia BH binaries, further orbital expansion during mass loss can lead to the development of a deep convective envelope in the donor. This may trigger a delayed CE phase. However, at this stage, the donor might have already lost a significant fraction of its envelope during the preceding mass loss phase. Consequently, the ejection of the remaining envelope could be facilitated.

If such a mass-loss mechanism operates, it could also influence the formation and predicted mass distributions of binary compact object mergers detected via gravitational waves, particularly those formed through isolated binary evolution involving mass-transfer interactions \citep{Belczynski2020, Bavera2021, Olejak2021a, vanSon2022, Picco2024, Olejak2024, Klencki2025}. A notable example of a potential gravitational-wave progenitor is the ultraluminous X-ray source P13 in NGC 7793 \citep{Bachetti2014, Motch2014, Furst2016}. This ULX binary consists of a massive late-B supergiant donor ($\sim 18$--$23\,M_\odot$) and a neutron star accretor ($\sim 1.4\,M_\odot$), exhibiting super-Eddington accretion. Despite the extreme mass ratio of $M_{\rm don}/M_{\rm acc} \approx 20$, the system appears to sustain stable mass transfer, implying reduced angular momentum loss, consistent with the scenario we propose for our Gaia BH progenitor systems with similar mass ratios. As discussed in Section~\ref{sec: mechanism}, our scenario may also be supported by observations of bipolar outflows surrounding some LBVs.

The next planned step is to develop and implement physical models within binary population synthesis frameworks to assess their broader impact on different populations. Incorporating this effect into rapid population synthesis properly is, however, challenging as it likely affects primarily a specific subset of systems—for example, those in particular mass, mass-ratio, orbital periods or metallicity regimes—depending on the underlying, still-speculative physical mechanism (or synergy of different factors). Future work should include high-resolution hydrodynamical simulations, more realistic modeling of the mass-loss stream that accounts, e.g., for self-accretion \citep{Hendriks2023}, combined with additional observational constraints from diverse binary populations, to test the plausibility and efficiency of the proposed mass-loss geometry. Upcoming Gaia data releases, in combination with complementary surveys, will be crucial for testing these models and further constraining the physical processes that govern binary evolution in the Galaxy and beyond.

%% Please use the acknowledgment and contribution environments. This will 
%% be anonomyized when the "anonymous" style option is used. 
\begin{acknowledgments}
A.O. would like to thank anonymous referee, Stephen Justham, Jim Fuller, Tassos Fragkos, Tomasz Bulik and Ruggero Valli for their comments. A.O. acknowledges useful discussions that inspired this work during the Stable Mass Transfer 2.0 Workshop at the Center for Computational Astrophysics in New York, organized with support from the Simons Foundation. This research was supported in part by grant NSF PHY-2309135 to the Kavli Institute for Theoretical Physics (KITP)

\end{acknowledgments}

\software{Python matplotlib, {\tt MESA}
          }

\appendix

\section{Appendix information} \label{Appendix}

\begin{figure} [ht]
    \centering
    \includegraphics[width=0.45\linewidth]{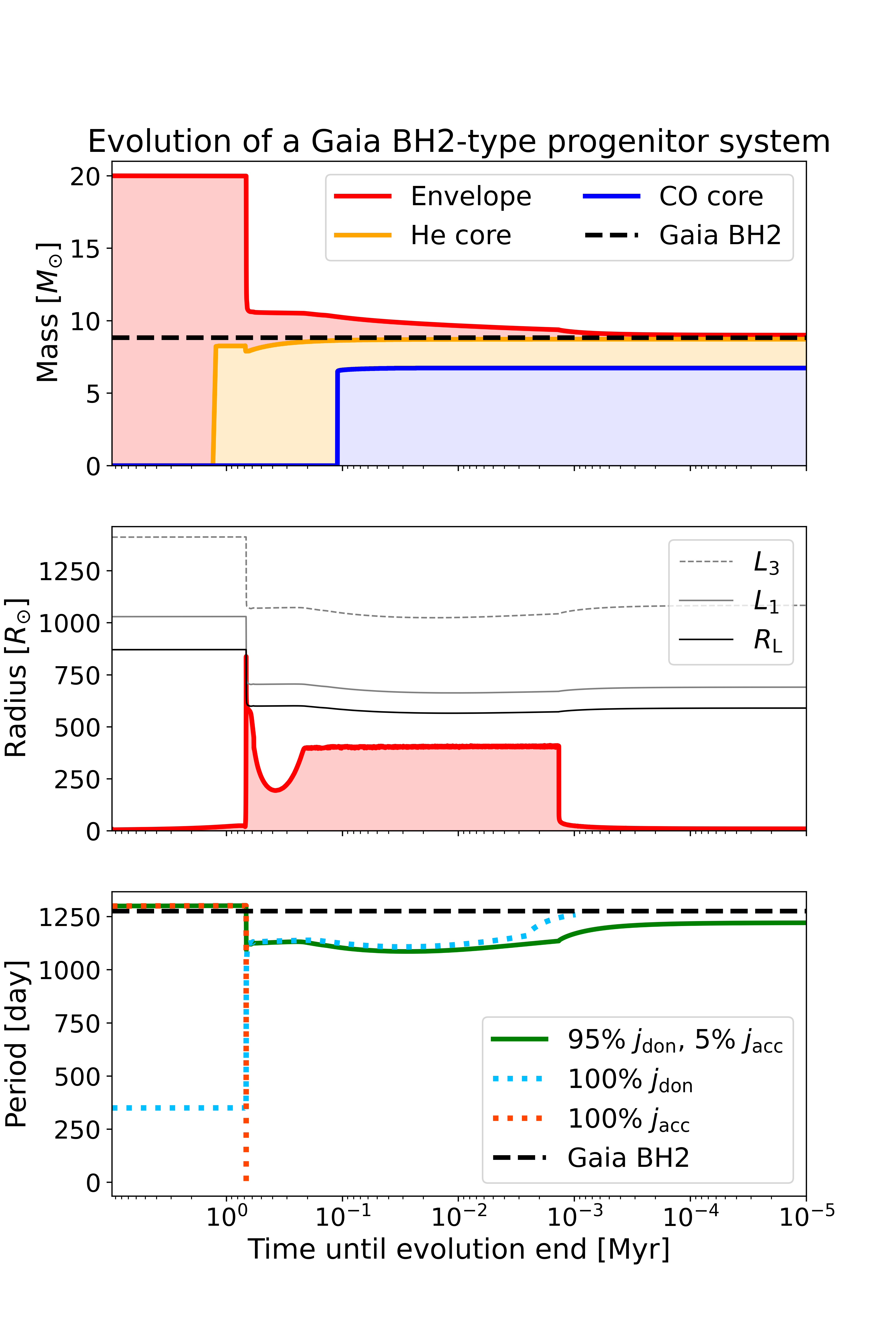}
    \caption{Evolution of the progenitor system of Gaia BH2. The physical assumption and figure description are the same as for Figure~\ref{Fig: Mass_evolution_1}.}
    \label{Fig: Mass_evolution_2}
\end{figure}

Figure~\ref{Fig: Mass_evolution_2} shows an example evolutionary track leading to a system with final properties (masses, orbital separation) equivalent to Gaia BH2. The top panel displays the evolution of the donor’s mass; the middle panel shows the evolution of the donor’s radius; and the bottom panel presents the evolution of the orbital period. The timescale is expressed as the logarithm of the remaining time until the end of evolution, defined as the point of central helium depletion.

The initial binary system consists of a $20\,{M}_\odot$ BH progenitor and a $1\,{M}_\odot$ companion in an orbit with a period of 1300 days, corresponding to a separation of $\sim 1383\,{R}_\odot$. After approximately $9$ Myr, the primary evolves off the main sequence and undergoes significant radial expansion, triggering a short but intense mass transfer episode lasting $\sim 0.1$ Myr, with a peak mass transfer rate of $3.8 \times 10^{-3}\,{M}_\odot\,\mathrm{yr}^{-1}$.

As a consequence of the mass transfer the donor rapidly loses nearly $10\,{M}_\odot$, stripping away most of its hydrogen-rich envelope. Following this, the donor’s radius contracts, and for over $\sim 0.4$ Myr the star detaches from its Roche lobe, halting mass transfer. At final stages of He burning, the star expands once more, without initiating a second mass transfer. The star loose additional an additional $\sim 1\,{M}_\odot$ in the stellar winds. Ultimately, the mass transfer ceases, leaving behind a $\sim 9.7\,{M}_\odot$ helium core and a low-mass companion in a widened orbit with a period of $\sim 1221$ days.

Within our default model, we assume that the mass lost from the system carries away 95\% of the specific angular momentum of the donor’s center of mass and 5\% of that of the accretor’s center of mass (green line in Fig.~\ref{Fig: Mass_evolution_2}), resulting in an approximately constant orbital separation. The orbital evolution is, however, highly sensitive to the adopted prescription for angular momentum loss.  For comparison, Fig.~\ref{Fig: Mass_evolution_2} also shows two limiting cases: (i) a model in which 100\% of the lost mass carries the specific angular momentum of the accretor, leading to rapid orbital shrinkage and likely stellar merger or common-envelope evolution (orange dotted line), and (ii) a model in which 100\% of the lost mass carries the specific angular momentum of the donor, resulting in continuous orbital widening (blue dotted line).

\section{Eccentricity} \label{sec: eccentricity}
Both Gaia BH1 and BH2 systems are eccentric (see Tab. \ref{tab: Gaia_parameters}). Formation of BHs with their final masses of $\sim 9-10\,{M_{\odot}}$ is highly uncertain. Since low mass BHs could form either via successful or failed supernova explosion - so-called direct collapse of the core \citep{Fryer2012,Ugliano2012,Burrows2020,Patton2020}, the eccentricity in the binaries may originate from multiple alternative mechanisms, which we briefly comment below:

\begin{itemize}
  \item \textit{Initial eccentricity:}  
  The progenitor binary could have formed with a non-zero eccentricity, which might be only partially damped if tidal circularization was inefficient. Both observations \citep{Oomen2018, Oomen2020, Escorza2020, Muller-Horn2025} and theoretical modeling \citep{Rocha2025, Parkosidis2025} suggests wide binaries can remain eccentric after mass transfer phase. 

  \item \textit{Asymmetric mass loss} may result in a natal kick imparted to the newly formed BH (tens of km s$^{-1}$) which efficiently induce orbital eccentricity \citep{Kalogera1996}. One can determine orbital parameters after the BH formation using equations \ref{eq. orbit} and \ref{eq. eccentricity}, for system separation and eccentricity, respectively \citep{Tauris2022}.
  
    \begin{equation} \label{eq. orbit}
    \frac{a_{\mathrm{post}}}{a_{\mathrm{pre}}} 
    = \left[
    2 - \frac{M_{\mathrm{pre}}}{M_{\mathrm{post}}}
    \left(
    \frac{v_r^2 + v_{\mathrm{p}}^2 + (V_{\mathrm{pre}} + v_{\mathrm{t}})^2}{V_{\mathrm{pre}}^2}
    \right)
    \right]^{-1}
    \end{equation}

    \begin{equation} \label{eq. eccentricity}
    1 - e^2 
    = \frac{M_{\mathrm{pre}}}{M_{\mathrm{post}}}
    \cdot \frac{a_{\mathrm{pre}}}{a_{\mathrm{post}}}
    \cdot
    \frac{v_{\mathrm{p}}^2 + (V_{\mathrm{pre}} + v_{\mathrm{t}})^2}{V_{\mathrm{pre}}^2}
    \end{equation}

    Where: $a_{\mathrm{pre}}$ and $a_{\mathrm{post}}$ are the semi-major axes before and after a BH formation, $e$ is the resulting eccentricity, $M_{\mathrm{pre}}$ and $M_{\mathrm{post}}$ are the total system masses before and after a BH formation, $V_{\mathrm{pre}}$ is the relative orbital velocity before a BH formation. The natal kick vector $v$ consists of components in the radial $v_{r}$), tangential $v_{t}$, and polar $v_{p}$ directions within the frame of reference of the exploding star.

    Since our model is flexible regarding the pre-supernova orbital parameters: separation and eccentricity, ejected masses and its direction, it is possible to reproduce the observed parameters for a variety of different initial assumptions. As an example, we can assume that the ejected mass to be $\Delta M = $ 0.5 M$_\odot$, consistent with a complete collapse scenario and mass loss only in the neutrino stream \citep[e.g.,][]{Fryer1999,Fryer2012,Janka2024} and that the system's orbit was circular before a BH formation.
  
    For a Gaia BH1 type of system, aiming to reach an eccentricity of $\approx$ 0.4 and a final orbital period of $\approx$ 180 days we can adopt for example: a radial kick of $\approx$ 35 km s$^{-1}$ with a pre-SN orbital period of $\approx$ 130 days or a tangential kick of $\approx$ 15 km s$^{-1}$ with a pre-SN orbital period of $\sim$ 90 days or a polar kick of $\approx$ 62 km s$^{-1}$ with a pre-SN orbital period of $\approx$ 85 days
    
    For Gaia BH2 type of system, targeting slightly higher an eccentricity of $\approx$ 0.5 and significantly wider orbital period of $\approx$ 1200 days we can find solutions: a radial kick of $\approx$ 25 kms$^{-1}$ with a pre-SN orbital period of $\approx$ 700 days, a tangential kick of $\approx$12 km s$^{-1}$ with a pre-SN orbital period of $\approx$ 450 days or a polar kick of $\approx$ 40 kms$^{-1}$ with a pre-SN orbital period of $\approx$ 450 days.
    
    Note that, to accommodate possible natal kicks, the progenitor systems shown in Figs. \ref{Fig: Mass_evolution_1} and \ref{Fig: Mass_evolution_2} would likely need to originate at somewhat smaller initial orbital separations, leaving sufficient parameter space for the kick-induced orbital changes. The numerical examples presented here are intended only to build intuition; in reality, the natal kick is expected to reflect a combination of these three variants. 

    \item \textit{Symmetric mass loss:}  
    Also, large spherically symmetric mass loss during BH formation can induce eccentricity via the so-called Blaauw kick, which can be estimated as \citep{Blaauw1961}:
    
    \begin{equation}
    e = \frac{M_{\mathrm{pre}} - M_{\mathrm{post}}}{M_{\mathrm{post}}}
    \end{equation}

  Where: ${M}_{\mathrm{pre}}$ is the total mass of the binary before the supernova, ${M}_{\mathrm{post}}$ is the total mass after the supernova.
  Reproducing the observed values ($e \approx 0.45$ for BH1 and $e \approx 0.52$ for BH2) would require substantial mass loss of $\sim$4–5 $M_{\odot}$.
  
  \item \textit{Dynamical interactions:}  
  Post-formation interactions—such as the presence of a distant tertiary companion—could increase the eccentricity through secular perturbations \citep[Kozai–Lidov cycles, see e.g., ][]{Shariat2023}. However, such scenarios require specific hierarchical triple configurations.

\end{itemize}

\bibliography{sample701}{}
\bibliographystyle{aasjournalv7}

\end{document}